\documentclass{article}
\usepackage{waspaa23,amsmath,graphicx,url,times}
\usepackage{color}

\usepackage[hidelinks,colorlinks=true]{hyperref} 
\usepackage{url}
\usepackage{color}
\usepackage{cite}
\usepackage{inconsolata}
\usepackage{amsmath}
\usepackage{amssymb}
\usepackage{booktabs}
\usepackage{microtype}
\usepackage{enumitem}
\usepackage[noabbrev,capitalise]{cleveref}
\usepackage{bm}
\usepackage{bbm}
\usepackage{multirow}
\usepackage{tabularx}

\usepackage{pifont}
\newcommand{\cmark}{\ding{51}}
\newcommand{\xmark}{\ding{55}}

\setlist{parsep=0ex,topsep=0.5ex,itemsep=0ex,leftmargin=1em}
\crefformat{footnote}{#2\footnotemark[#1]#3}
\crefrangeformat{footnote}{#3\footnotemark[#1]#4--#5\footnotemark[#2]#6}
\crefmultiformat{footnote}{#2\footnotemark[#1]#3}{\textsuperscript{,}#2\footnotemark[#1]#3}{\textsuperscript{,}#2\footnotemark[#1]#3}{\textsuperscript{,}#2\footnotemark[#1]#3}
\graphicspath{{figs/}}

\newcolumntype{C}{>{\centering\arraybackslash}X}
\definecolor{darkred}{rgb}{0.6,0.0,0.0}
\definecolor{darkblue}{rgb}{0.0,0.0,0.75}
\definecolor{myblue}{rgb}{0,0.3,0.6}
\hypersetup{linkcolor=myblue,urlcolor=myblue,citecolor=myblue,anchorcolor=myblue}

\title{CLIPSonic: Text-to-Audio Synthesis with Unlabeled Videos\\and Pretrained Language-Vision Models}

\name{\shortstack{Hao-Wen~Dong$^{1,2}$\sthanks{Work done during an internship at Dolby. Hao-Wen thanks Taiwan Ministry of Education for supporting his PhD study. Contact: \texttt{\href{mailto:hwdong@ucsd.edu}{hwdong@ucsd.edu}}}\quad Xiaoyu~Liu$^1$\quad Jordi~Pons$^1$\quad Gautam~Bhattacharya$^1$\\\itshape Santiago~Pascual$^1$\quad Joan~Serrà$^1$\quad Taylor~Berg-Kirkpatrick$^2$\quad Julian~McAuley$^2$}\vspace{-1.05ex}}
\address{$^1$\,Dolby Laboratories\quad$^2$\,University of California San Diego}

\begin{document}

\ninept
\maketitle

\sloppy

\begin{abstract}
Recent work has studied text-to-audio synthesis using large amounts of paired text-audio data. However, audio recordings with high-quality text annotations can be difficult to acquire. In this work, we approach text-to-audio synthesis using unlabeled videos and pretrained language-vision models. We propose to learn the desired text-audio correspondence by leveraging the visual modality as a bridge. We train a conditional diffusion model to generate the audio track of a video, given a video frame encoded by a pretrained contrastive language-image pretraining (CLIP) model. At test time, we first explore performing a zero-shot modality transfer and condition the diffusion model with a CLIP-encoded text query. However, we observe a noticeable performance drop with respect to image queries. To close this gap, we further adopt a pretrained diffusion prior model to generate a CLIP image embedding given a CLIP text embedding. Our results show the effectiveness of the proposed method, and that the pretrained diffusion prior can reduce the modality transfer gap. While we focus on text-to-audio synthesis, the proposed model can also generate audio from image queries, and it shows competitive performance against a state-of-the-art image-to-audio synthesis model in a subjective listening test. This study offers a new direction of approaching text-to-audio synthesis that leverages the naturally-occurring audio-visual correspondence in videos and the power of pretrained language-vision models.
\end{abstract}


\section{Introduction}
\label{sec:intro}

\begin{figure}[t]
    \centering
    \includegraphics[width=0.99\linewidth]{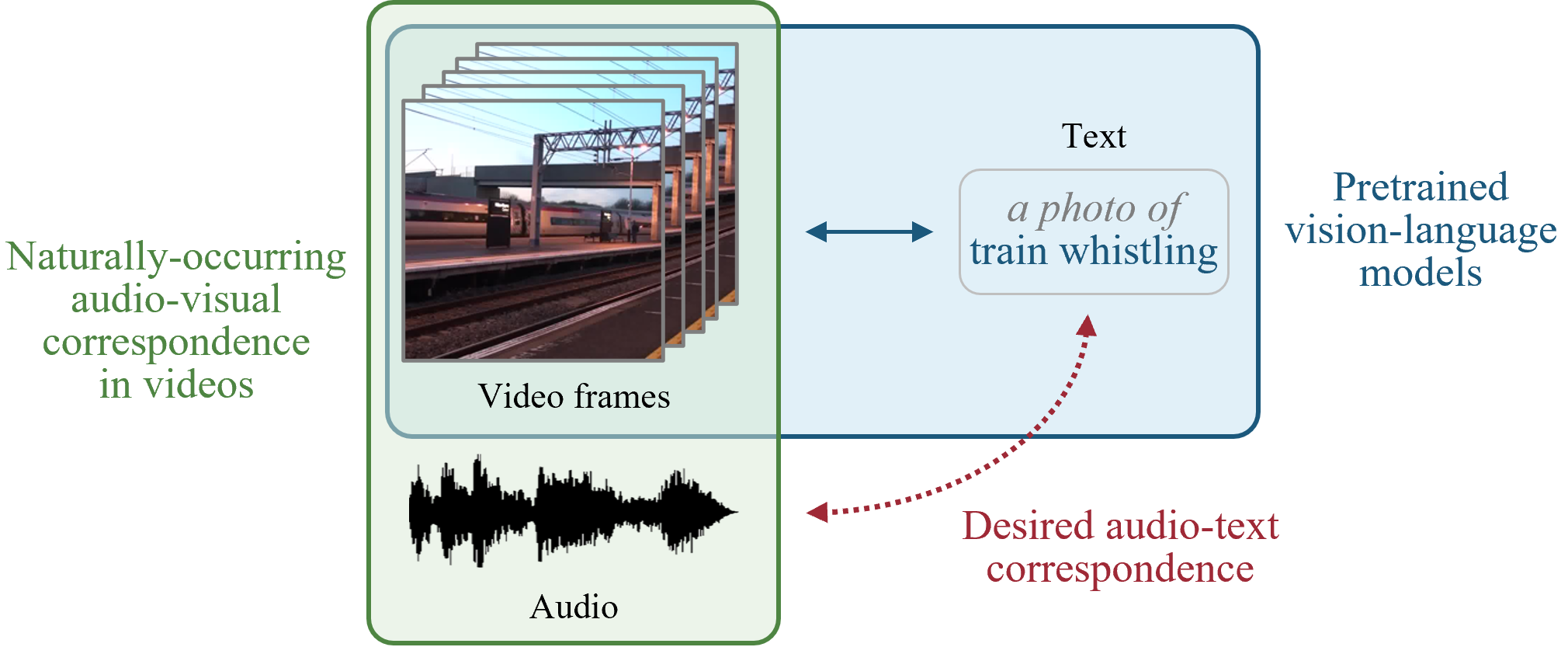}\vspace{-2ex}
    \caption{We learn the text-audio correspondence by leveraging the audio-visual correspondences in videos and the multimodal representation learned by pretrained language-vision models.}
    \vspace{-2ex}
    \label{fig:bridge}
\end{figure}

With the advance of generative modeling~\cite{radford2019gpt2,ho2020ddpm,rombach2022ldm} and language-audio contrastive learning~\cite{wu2022clap,huang2022mulan,guzhov2022audioclip}, various deep learning-based text-to-audio synthesis systems have recently emerged~\cite{yang2022diffsound,kreuk2022audiogen,liu2023audioldm,huang2023makeanaudio,huang2023noise2music,agostinelli2023musiclm}. However, these systems typically require a large amount of paired text-audio data for training. Despite extensive human annotation efforts, the current largest public text-audio dataset contains around 630\,k text-audio pairs~\cite{wu2022clap}. Given the relative scarcity of text-audio data on the web as compared to text-image data, it remains unclear whether we can scale up text-audio datasets to a size comparable with large scale text-image datasets, e.g., the LAION-5B dataset~\cite{schuhmann2022laion5b}, which contains 5.85 billion text-image pairs. In this work, we approach text-to-audio synthesis without text-audio pairs through leveraging the naturally-occurring audio-visual correspondence in videos and the multimodal representation learned by pretrained language-vision models (see~\cref{fig:bridge}).

\begin{figure*}
    \centering
    \includegraphics[width=0.96\linewidth]{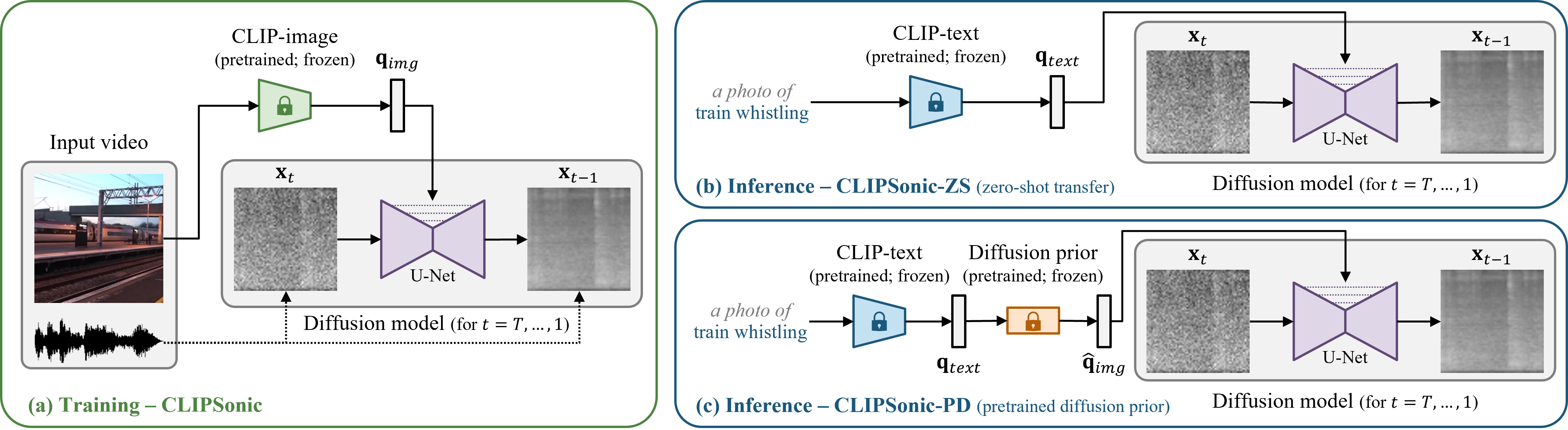}\vspace{-2ex}
    \caption{Proposed CLIPSonic model. During training, CLIPSonic learns to synthesize the audio track of a video given the image in a video frame. At inference time, we feed a text query in the form of ``a photo of [label]'' to approach text-to-audio synthesis or use a pretrained diffusion prior model to close the gap between  the text queries (used for inference) and the image queries (used for training). $\textbf{x}_t$ represents a noisy spectrogram at diffusion step $t$. The generated mel spectrogram $\hat{\textbf{x}}_0$ is inverted back to waveform by a pretrained BigVGAN model~\cite{lee2023bigvgan}.}
    \vspace{-1ex}
    \label{fig:model}
\end{figure*}

The proposed CLIPSonic model is based on a conditional diffusion model~\cite{nichol2021ddpm}, a constrastive language-image pretraining (CLIP) model~\cite{radford2021clip}, and a pretrained diffusion prior model~\cite{ramesh2022dalle2}, as illustrated in \cref{fig:model}. Given a video, CLIPSonic is trained to synthesize the mel spectrogram of the audio given a CLIP-encoded frame, randomly selected from the video. Since CLIP embeds images and texts into a cross-modal semantic space, CLIPSonic learns to map the CLIP embedding space to audio. At test time, we first explore performing a zero-shot modality transfer and conditioning the diffusion model directly with a CLIP-encoded text query. However, we observe in practice a noticeable performance drop with respect to image queries. To close this gap, we adopt a pretrained diffusion prior model to generate a CLIP image embedding given a CLIP text embedding. We note that our proposed system requires only 1) unlabeled videos, for training the conditional diffusion model, and 2) image-text pairs, for pre-training the language-vision models. Through a subjective listening test and an objective evaluation, our experimental results demonstrate the effectiveness of the proposed method. Audio samples are available on our demo website.\footnote{\url{https://salu133445.github.io/clipsonic/}\label{fn:demo}}

Our study differs from prior work in several ways. Existing text-to-audio models rely on large amounts of text-audio training pairs~\cite{yang2022diffsound,kreuk2022audiogen,liu2023audioldm,huang2023makeanaudio,huang2023noise2music,agostinelli2023musiclm}, 
whereas CLIPSonic
learns text-queried audio synthesis without text-audio pairs. Prior work studied image-to-audio synthesis~\cite{owens2016vis,iashin2021specvqgan,sheffer2022im2wav}, but they do not examine the zero-shot modality transfer between texts and images. CLIPSep~\cite{dong2023clipsep} and CLIPSynth~\cite{ hdong2023clipsynth} propose to learn text-queried source separation and audio synthesis from unlabeled videos, respectively, but they do not address the issue of the zero-shot modality transfer gap. DALL-E 2~\cite{ramesh2022dalle2} proposes the diffusion prior model to address the zero-shot modality transfer gap in CLIP-based text-to-image synthesis, and we explore leveraging a pretrained diffusion prior model to transfer the knowledge learned from videos for text-to-audio synthesis. Other related works are AudioLDM~\cite{liu2023audioldm} and MusicLM~\cite{agostinelli2023musiclm}, which rely on language-audio models~\cite{wu2022clap,huang2022mulan} to perform a zero-shot audio-to-text modality transfer, but such language-audio models are trained on audio-text pairs.

\section{CLIPSonic}
\label{sec:method}

In this section, we introduce the proposed CLIPSonic model for learning text-to-audio synthesis from unlabeled videos. As illustrated in \cref{fig:model}(a), CLIPSonic uses a mel spectrogram-based diffusion model for audio synthesis. We adopt the diffusion framework for its strong performance in audio synthesis~\cite{kong2021diffwave,pascual2022diffusion,liu2023audioldm}. Given a video, CLIPSonic is trained to synthesize the mel spectrogram of the audio from the image in a randomly extracted video frame. Specifically, we first use a pretrained CLIP image encoder to encode the image into a query vector $\mathbf{q}_\mathit{img}$. Then, this query vector is used as a conditional signal to guide the diffusion model to generate a mel spectrogram $\hat{\mathbf{x}}_0$. We adopt a denoising diffusion probabilistic model~\cite{nichol2021ddpm} and classifier-free guidance~\cite{ho2021classifierfree}, which allows us to control the degree of conditioning signal through the guidance level variable $w$ during inference.\footnote{We use the formulation: $\nabla_\mathbf{x} \log p_w (\mathbf{x}\,|\,\mathbf{q}) = (1 - w) \nabla_\mathbf{x} \log p(\mathbf{x}) + w \nabla_\mathbf{x} \log p(\mathbf{x}\,|\,\mathbf{q})$. A larger $w$ leads to a stronger conditioning signal, and $w = 1$ corresponds to a conditional model without classifier-free guidance.}
The generated mel spectrograms are inverted back to waveforms using a separately-trained BigVGAN~\cite{lee2023bigvgan}. We choose to perform diffusion on the mel spectrogram domain for its lower dimensionality than waveforms, and because BigVGAN shows good quality when synthesizing general audio from mel spectrograms.

\vspace{1ex}

\noindent\textbf{CLIPSonic-ZS (zero-shot modality transfer).}\quad
At inference time, we aim to leverage the language-vision embedding space learned by CLIP to achieve text-to-audio synthesis. CLIPSonic-ZS explores swapping the CLIP image embeddings for the CLIP text embeddings, as a way to use text queries in a zero-shot modality transfer setting. As illustrated in \cref{fig:model}(b), we use the CLIP text encoder to encode the input text query into a query vector $\mathbf{q}_\mathit{text}$, which is fed as a condition to the diffusion model. We refer to this model as CLIPSonic-ZS, where ``ZS'' stands for \underline{z}ero-\underline{s}hot modality transfer.

\vspace{1ex}

\noindent\textbf{CLIPSonic-PD (pretrained diffusion prior).}\quad
As to be shown in \cref{sec:results}, we observe a modality gap between CLIP's text and image embedding spaces. Following DALL-E 2~\cite{ramesh2022dalle2}, we explore relying on a diffusion prior model to bridge this gap. As illustrated in \cref{fig:model}(c), we first encode the input text query into a CLIP text embedding vector $\mathbf{q}_\mathit{text}$ and then generate a CLIP image embedding vector $\hat{\mathbf{q}}_\mathit{img}$ from $\mathbf{q}_\mathit{text}$ using the pretrained diffusion prior model. The generated query vector $\hat{\mathbf{q}}_\mathit{img}$ is then passed as the conditioning signal to the diffusion model. We refer to this model as \mbox{CLIPSonic-PD} (\underline{p}retrained \underline{d}iffusion prior). Note that both CLIPSonic-ZS and CLIPSonic-PD require no text-audio pairs for training. Further, both the CLIP and diffusion prior models can be pretrained using only text-image pairs, hence suppressing the need for paired audio-text data.

\vspace{1ex}

\noindent\textbf{CLIPSonic-IQ and CLIPSonic-SD.}\quad
While here we focus on text-to-audio, CLIPSonic can also be used as an image-to-audio synthesis model by using ${\mathbf{q}}_\mathit{img}$ queries. We will refer to this variant as CLIPSonic-IQ (\underline{i}mage-\underline{q}ueried). Moreover, we find that it is possible to train the diffusion prior model from scratch on domain-specific datasets, and hence we also consider a variant called CLIPSonic-SD (\underline{s}upervised \underline{d}iffusion prior), where we train the diffusion prior model from scratch using text-image pairs in our datasets. As will be specified in \cref{sec:setup}, since the text data used to train the diffusion prior in CLIPSonic-SD comes from audio labels in this work, CLIPSonic-SD serves as an oracle model against CLIPSonic-PD. By comparing CLIPSonic-PD to CLIPSonic-SD, we intend to study the effectiveness of using a diffusion prior model pretrained on a massive amount of data against one trained on the target dataset.

\begin{figure*}[t]
    \centering
    \includegraphics[width=\linewidth]{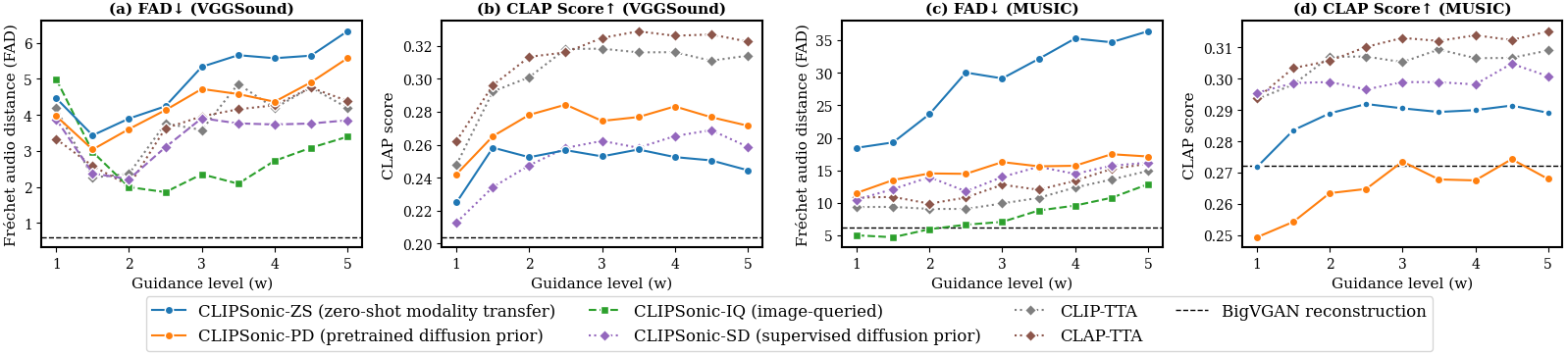}
    \vspace{-5ex}
    \caption{Objective evaluation results on VGGSound and MUSIC.}
    \vspace{-1ex}
    \label{fig:results}
\end{figure*}

\section{Experimental setup}
\label{sec:setup}

\textbf{Data.}\quad
We consider two datasets: VGGSound~\cite{chen2020vggsound} and MUSIC~\cite{zhao2018sop}. The VGGSound dataset consists of 171,899 10-sec YouTube videos, covering 310 classes of sounds in the wild, and we follow the train-test split provided with the dataset. The MUSIC dataset consists of 1,055 full-length YouTube videos of people playing a musical instrument, with 21 instrument types in total. We randomly split the dataset into a 9:1 train-test split.
VGGSound represents a large, diverse dataset captured from unstructured sources in the wild, whereas MUSIC represents a small, curated dataset of a specific domain of interest. As both datasets come with only class labels, we convert such labels \mbox{into pseudo text in the form of ``a photo of [label]''.}

\vspace{1ex}

\noindent\textbf{Baseline models.}\quad
We compare CLIPSonic models against the following text-to-audio (TTA) and reconstruction models. 
\begin{itemize}
    \item \textbf{CLIP-TTA} is the supervised version of CLIPSonic where we use text-audio pairs for training. The pretrained CLIP-text embedding is used as conditioning.
    \item \textbf{CLAP-TTA} is the same as CLIP-TTA but uses pretrained CLAP-text embeddings~\cite{wu2022clap}, where we use a prompt in the form of ``the sound of [label]''. Unlike CLIP-text embeddings, CLAP-text embeddings are expected to encode audio-grounded features rather than visually-grounded features.
    \item \textbf{BigVGAN mel spectrogram reconstruction} are waveforms reconstructed from the ground-truth mel spectrograms by the BigVGAN model. This serves as an upper bound of spectrogram-based synthesis systems that use BigVGAN as the inversion model.
\end{itemize}

\vspace{1ex}

\noindent\textbf{Implementation details.}\quad
For mel spectrogram computation, we use a sampling rate of 16\,kHz, a hop size of 512, an FFT filter size of 2048, and 64~mel bands. During training, we use mel spectrograms of size 64$\times$64, which corresponds to two seconds of audio. For the diffusion model, we follow the network architecture proposed in~\cite{nichol2021ddpm} and use the open-source code in~\cite{improved_diff_github}. We use a cosine noise schedule with 4000 diffusion steps during training and 1000 steps at inference time. We use AdamW with a learning rate of 0.0001, a batch size of 32, and a dropout rate of 0.1 in classifier-free guidance. All diffusion models are trained for 200\,k steps on MUSIC and 500\,k steps on VGGSound using two NVIDIA RTX 2080 Ti GPUs, which takes a day on MUSIC and two days on VGGSound. For the pretrained CLIP model, we use the ``ViT-L/14'' version trained on 400~million image-text pairs~\cite{clip_github}. We use a pretrained transformer-based diffusion prior model trained on 2~billion image-text pairs using the same backbone CLIP model~\cite{dalle2_huggingface}. For training the diffusion prior model CLIPSonic-SD from scratch, we follow the same architecture as in CLIPSonic-PD and use the code in~\cite{dalle2_github}. We use AdamW with a learning rate of 0.0001 and a batch size of 32. The diffusion prior models are trained on MUSIC and VGGSound, respectively, until convergence at around 200\,k steps, which takes a day on a NVIDIA RTX 2080 Ti GPU. For the CLAP model, we use the ``630k-audioset-fusion'' version released in~\cite{clap_github}. For the BigVGAN model, we pretrain it on VGGSound for 500\,k steps using the code in~\cite{bigvgan_github} and use this pretrained version in all of our experiments.

\vspace{1ex}

\noindent\textbf{Evaluation metrics.}\quad
To compare the performance of our method against the baselines, we sample 512 audio samples from each model and compute the Fréchet audio distance (FAD)~\cite{kilgour2019fad} and the CLAP score~\cite{huang2023makeanaudio, wu2022clap}. The FAD measures how close the generated audio samples are to the reference audio in terms of quality and diversity.\footnote{Following~\cite{yang2022diffsound,liu2023audioldm} we also computed the Fréchet inception distance (FID) of the generated spectrograms, and found that the trend of FID aligned well with that of FAD. For brevity, we only report and discuss the FAD results.} We adopt the open-source implementation provided in~\cite{fad_github} and use the VGGish~\cite{hershey2017cnn} as the backbone model for FAD. The CLAP score measures the relevance between the generated audio and the input query text, and it is formally defined as the cosine similarity between the CLAP embedding of the audio and that of the input text query.

\vspace{1ex}

\noindent\textbf{Subjective test.}\quad
We conduct a listening test to study the fidelity of the generated audio and their relevance to textual (text-to-audio) and visual (image-to-audio) prompts. We ask 21 expert listeners to rate the generated audio samples on a 1--5 scale in terms of fidelity and relevance. Fidelity experiments study the quality of the generated audio (without evaluating its semantic grounding) while relevance experiments study the semantic correspondence with respect to the prompt (without evaluating its audio quality). The audio samples used for this test are available on our demo website.\cref{fn:demo}

\section{Results}
\label{sec:results}

\subsection{Objective Evaluation Results}
\label{sec:obj}

\noindent\textbf{Guidance.}\quad
\cref{fig:results} shows the results of the studied models as a function of the classifier-free guidance scale $w$. As noted in \cite{ho2021classifierfree} using conceptually-similar measures, the different curves between FAD and CLAP scores imply a trade-off between quality/diversity (represented by FAD) and query-sample relevance (represented by CLAP score). Noticeably, in terms of CLAP score, all models (except CLIPSonic-PD on MUSIC) outperform the BigVGAN reconstruction on both datasets (see \cref{fig:results}(b) and (d)). We attribute the higher CLAP scores to the classifier-free guidance as it is shown to improve adherence to the conditioning \cite{ho2021classifierfree} but at the cost of diversity---note the increasing FAD in Figure \ref{fig:results}(a) and (c) as $w$ increases. As such, practitioners can choose $w$ based on their specific requirements. We use $w = 1.5$ as it offers a good balance between quality/diversity and relevance, and we report the results in \cref{tab:results}.

\begin{table*}[t]
    \centering
    \caption{Evaluation results on VGGSound and MUSIC datasets, evaluated at $w = 1.5$.}
    \vspace{1ex}
    \resizebox{\linewidth}{!}{
    \begin{tabular}{lcc@{~~~~}cc@{~~~~}cc@{~~~~}c}
        \toprule
        \multirow{2}{*}[-3pt]{Model} &\multirow{2}{*}[-3pt]{\shortstack{Without\\text-audio pairs}} &\multicolumn{2}{c}{Query modality} &\multicolumn{2}{c}{VGGSound} &\multicolumn{2}{c}{MUSIC}\\
        \cmidrule(lr){3-4} \cmidrule(lr){5-6} \cmidrule(lr){7-8}
        & & Training & Inference &FAD $\downarrow$ &CLAP score $\uparrow$ &FAD $\downarrow$  &CLAP score $\uparrow$\\
        \midrule
        CLIPSonic-IQ {\footnotesize(image-queried)}               &-      &Image &Image &2.97 &-     &4.71  &-\\
        CLIPSonic-ZS {\footnotesize(zero-shot modality transfer)} &\cmark &Image &Text  &3.43 &0.258 &19.30 &0.284 \\
        CLIPSonic-PD {\footnotesize(pretrained diffusion prior)}  &\cmark &Image &Text  &3.04 &0.265 &13.51 &0.254 \\
        \midrule
        CLIPSonic-SD {\footnotesize(supervised diffusion prior)}  &\xmark &Image &Text  &2.37 &0.234 &12.13 &0.299\\
        CLIP-TTA                                                  &\xmark &Text  &Text  &2.26 &0.292 &9.39  &0.298\\
        CLAP-TTA                                                  &\xmark &Text  &Text  &2.58 &0.296 &10.92 &0.303\\
        \midrule
        BigVGAN mel spectrogram reconstruction                    &-      &-     &-     &0.60 &0.204 &6.21  &0.272\\
        \bottomrule
    \end{tabular}
    }
    \vspace{-1ex}
    \label{tab:results}
\end{table*}

\vspace{1ex}

\noindent\textbf{Models without text-audio pairs.}\quad
First, we discuss CLIPSonic models in \cref{tab:results} that do not use text-audio pairs during training. CLIPSonic-IQ (image-queried) achieves a strong performance on both datasets. Yet, when we switch to using text queries in a zero-shot setting with CLIPSonic-ZS, we observe a performance drop in terms of FAD on both datasets. This performance drop suggests a modality gap between CLIP's image (used during training) and text (used during inference) embedding spaces. In contrast, with the pretrained diffusion prior model, CLIPSonic-PD achieves a lower FAD than CLIPSonic-ZS across different $w$ values (see also \cref{fig:results}). To further investigate this, we report in \cref{tab:cosine_sim} the average cosine similarity between the query embedding ($\mathbf{q}_\mathit{text}$ or $\hat{\mathbf{q}}_\mathit{img}$) and the ground truth CLIP-image embedding $\mathbf{q}_\mathit{img}$. We note that CLIPSonic-ZS leads to a low cosine similarity, which supports our hypothesis that there is a modality gap in CLIP's embedding space. In contrast, CLIPSonic-PD achieves a significantly higher cosine similarity, showing that the pretrained diffusion prior model can effectively bridge the modality gap. Moreover, while we observe a lower CLAP score for CLIPSonic-PD on MUSIC, we observe little difference in the relevance criterion in the listening test to be discussed in \cref{sec:subjective} (see \cref{tab:subjective}), suggesting that all these models have passed a reasonable level of audio-text relevance.

\begin{table}[t]
    \centering
    \vspace{-3ex}
    \caption{Cosine similarities between various query embeddings.}
    \vspace{1ex}
    \begin{tabular}{lc@{~~}c@{~~~}c}
        \toprule
        Model        &Similarity computed &VGGSound &MUSIC \\ 
        \midrule
        CLIPSonic-ZS &$\mathrm{sim}(\mathbf{q}_\mathit{text}, \mathbf{q}_\mathit{img})$ &0.205 &0.245\\
        CLIPSonic-PD &$\mathrm{sim}(\hat{\mathbf{q}}_\mathit{img}, \mathbf{q}_\mathit{img})$ &0.647 &0.720\\
        CLIPSonic-SD &$\mathrm{sim}(\hat{\mathbf{q}}_\mathit{img}, \mathbf{q}_\mathit{img})$ &0.711 &0.824\\
        \bottomrule
    \end{tabular}
    \label{tab:cosine_sim}
\end{table}

\vspace{1ex}

\noindent\textbf{Models using text-audio pairs.}\quad
We now compare the baseline models that do use text-audio pairs for training against the previous CLIPSonic variants. 
First, we see that CLIPSonic-SD, with a diffusion prior trained directly on the target dataset, achieves a lower FAD than CLIPSonic-PD, which uses the pretrained diffusion prior. This is possibly due to the distribution mismatch between the target datasets and the LAION-2B dataset used to train the pretrained prior.\footnote{We note that there is also a mismatch in the semantics of the textual queries, where the target datasets contain audio-specific labels while LAION-2B contains visually-grounded labels. However, the similar performance of CLIP-TTA and CLAP-TTA suggests that this is a minor effect.} From \cref{tab:cosine_sim}, we can also see that CLIPSonic-SD can generate a CLIP-image embedding closer to the ground truth embedding on the target datasets than CLIPSonic-PD. Yet, in our subjective evaluation below we will see that CLIPSonic-PD still exhibits a favorable degree of generalization to downstream datasets since it consistently outperforms CLIPSonic-ZS. Moreover, we observe a gap between the performance of CLIPSonic-PD and that of CLIP-TTA and CLAP-TTA. However, we note that this is an unfair comparison as CLIP-TTA and CLAP-TTA are trained on audio-text pairs, while CLIPSonic-PD does not use audio-text pairs in training.

\subsection{Subjective Listening Test Results}
\label{sec:subjective}

\begin{table}[t]
    \centering
    \vspace{-2.5ex}
    \caption{Listening test results for text-to-audio synthesis (MOS).}
    \vspace{1ex}
    \resizebox{\linewidth}{!}{
    \begin{tabular}{l@{~~}c@{~~}c@{~~}c@{~~}c}
        \toprule
        \multirow{2}{*}[-.5ex]{Model} &\multicolumn{2}{c}{VGGSound} &\multicolumn{2}{c}{MUSIC}\\
        \cmidrule(lr){2-3} \cmidrule(lr){4-5}
        &Fidelity &Relevance &Fidelity &Relevance\\
        \midrule
        CLIPSonic-ZS &{2.55 $\pm$ 0.22} &{2.01 $\pm$ 0.27} &{2.98 $\pm$ 0.23} &{3.87 $\pm$ 0.24}\\
        CLIPSonic-PD &\textbf{3.04 $\pm$ 0.20} &{2.86 $\pm$ 0.25} &\textbf{3.67 $\pm$ 0.18} &{3.91 $\pm$ 0.24}\\
        CLIPSonic-SD &{2.96 $\pm$ 0.21} &\textbf{3.49 $\pm$ 0.28} &{3.36 $\pm$ 0.20} &\textbf{4.07 $\pm$ 0.22}\\
        \cmidrule(lr){1-5}
        Ground truth &{3.78 $\pm$ 0.19} &{3.54 $\pm$ 0.29} &{3.90 $\pm$ 0.17} &{4.34 $\pm$ 0.18}\\
        \bottomrule
    \end{tabular}
    }
    \vspace{-2ex}
    \label{tab:subjective}
\end{table}

\begin{table}[t]
    \centering
    \vspace{-3ex}
    \caption{Listening test results for image-to-audio synthesis (MOS).}
    \vspace{1ex}
    \begin{tabular}{lcc}
        \toprule
        Model &Fidelity &Relevance\\
        \midrule
        CLIPSonic-IQ (image-queried)         &\textbf{3.29 $\pm$ 0.16} &3.80 $\pm$ 0.19\\
        SpecVQGAN \cite{iashin2021specvqgan} &2.15 $\pm$ 0.17 & 2.54 $\pm$ 0.23\\
        \textsc{Im2Wav} \cite{sheffer2022im2wav}      &2.19 $\pm$ 0.15 & \textbf{3.90 $\pm$ 0.22}\\
        \bottomrule
    \end{tabular}
    \vspace{-2.8ex}
    \label{tab:subjective_image}
\end{table}

\noindent\textbf{Text-to-audio synthesis.}\quad
We conduct an ablation study to compare CLIPSonic-ZS, -PD and -SD variants on MUSIC and VGGSound. As shown in \cref{tab:subjective}, CLIPSonic-ZS consistently underperforms, arguably because of the aforementioned mismatch between text and image embeddings. The two contributed variants, i.e., CLIPSonic-PD and -SD, consistently achieve higher MOS than CLIPSonic-ZS, both in terms of relevance and fidelity. Notably, the ground truth scores are relatively low (an MOS between 3 to 4), especially noticeable for VGGSound as it is noisier than the MUSIC dataset.

\vspace{1ex}

\noindent\textbf{Image-to-audio synthesis.}\quad
While our focus is to study text-to-audio synthesis, CLIPSonic-IQ can also generate audio from image queries. We compare it against SpecVQGAN~\cite{iashin2021specvqgan}, a representative image-to-audio model, and \textsc{Im2Wav}~\cite{sheffer2022im2wav}, a state-of-the-art model for image-to-audio synthesis. 
All three models are trained on VGGSound and tested on out-of-distribution samples from \textsc{ImageHear}~\cite{sheffer2022im2wav}. The selected samples conform a challenging benchmark for us because they are 1) selected from \textsc{Im2Wav}'s demo website and 2) out-of-distribution. As shown in \cref{tab:subjective_image}, CLIPSonic-IQ outperforms the state-of-the-art in fidelity, while remaining competitive in terms of relevance. The improved fidelity can possibly be attributed to the fact that we use a continuous representation (mel spectrogram) with a state-of-the-art inversion model (BigVGAN), as compared to the discrete VQ-VAE representation used in \textsc{Im2Wav}.

\section{Conclusion}
\label{sec:conclusion}

We explored approaching text-to-audio synthesis without text-audio pairs by using unlabeled videos and pretrained language-vision models. Through both objective and subjective evaluations, we showed that the proposed models can effectively learn text-to-audio synthesis without text-audio pairs, and the pretrained diffusion prior can reduce the modality transfer gap caused by the mismatch between CLIP’s image (used for training) and text (used for inference) embedding spaces. Moreover, in a subjective listening test, the image-to-audio synthesis model that we base our modality transfer upon achieves competitive performance against a state-of-the-art image-to-audio synthesis model. Finally, we argue that images provide rich conditioning signals for audio synthesis, and leveraging such rich signals to improve text-to-audio synthesis is a promising research direction. Along this direction, CLIPSonic represents an example using videos and pretrained language-vision models. For future work, we intend to scale up the proposed method to a larger amount of videos, and explore using tri-modal audio-vision-language models~\cite{guzhov2022audioclip,wu2022wav2clip,rouditchenko2021avlnet}.

\clearpage
\bibliographystyle{IEEEtran}
\bibliography{ref}

\begin{thebibliography}{10}
\providecommand{\url}[1]{#1}
\def\UrlFont{\rmfamily}
\providecommand{\newblock}{\relax}
\providecommand{\bibinfo}[2]{#2}
\providecommand\BIBentrySTDinterwordspacing{\spaceskip=0pt\relax}
\providecommand\BIBentryALTinterwordstretchfactor{4}
\providecommand\BIBentryALTinterwordspacing{\spaceskip=\fontdimen2\font plus
\BIBentryALTinterwordstretchfactor\fontdimen3\font minus
  \fontdimen4\font\relax}
\providecommand\BIBforeignlanguage[2]{{%
\expandafter\ifx\csname l@#1\endcsname\relax
\typeout{** WARNING: IEEEtran.bst: No hyphenation pattern has been}%
\typeout{** loaded for the language `#1'. Using the pattern for}%
\typeout{** the default language instead.}%
\else
\language=\csname l@#1\endcsname
\fi
#2}}

\bibitem{radford2019gpt2}
A.~Radford, J.~Wu, R.~Child, D.~Luan, D.~Amodei, and I.~Sutskever, ``{Language
  Models are Unsupervised Multitask Learners},'' \emph{Technical Report of
  OpenAI}, 2019.

\bibitem{ho2020ddpm}
J.~Ho, A.~Jain, and P.~Abbeel, ``{Denoising Diffusion Probabilistic Models},''
  in \emph{Proc. NeurIPS}, 2020.

\bibitem{rombach2022ldm}
R.~Rombach, A.~Blattmann, D.~Lorenz, P.~Esser, and B.~Ommer, ``{High-resolution
  image synthesis with latent diffusion models},'' in \emph{Proc. CVPR}, 2022,
  pp. 10\,684--10\,695.

\bibitem{wu2022clap}
Y.~Wu, K.~Chen, T.~Zhang, Y.~Hui, T.~Berg-Kirkpatrick, and S.~Dubnov,
  ``{Large-scale Contrastive Language-Audio Pretraining with Feature Fusion and
  Keyword-to-Caption Augmentation},'' in \emph{Proc. ICASSP}, 2023.

\bibitem{huang2022mulan}
Q.~Huang, A.~Jansen, J.~Lee, R.~Ganti, J.~Y. Li, and D.~P.~W. Ellis,
  ``{{M}u{L}an: A Joint Embedding of Music Audio and Natural Language},'' in
  \emph{Proc. ISMIR}, 2022.

\bibitem{guzhov2022audioclip}
A.~Guzhov, F.~Raue, J.~Hees, and A.~Dengel, ``{Audio{CLIP}: Extending {CLIP} to
  Image, Text and Audio},'' in \emph{Proc. ICASSP}, 2022, pp. 976--980.

\bibitem{yang2022diffsound}
D.~Yang, J.~Yu, H.~Wang, W.~Wang, C.~Weng, Y.~Zou, and D.~Yu, ``{Diffsound:
  Discrete Diffusion Model for Text-to-sound Generation},'' \emph{arXiv
  preprint arXiv:2207.09983}, 2022.

\bibitem{kreuk2022audiogen}
F.~Kreuk, G.~Synnaeve, A.~Polyak, U.~Singer, A.~Défossez, J.~Copet, D.~Parikh,
  Y.~Taigman, and Y.~Adi, ``{{A}udio{G}en: Textually Guided Audio
  Generation},'' in \emph{Proc. ICLR}, 2023.

\bibitem{liu2023audioldm}
H.~Liu, Z.~Chen, Y.~Yuan, X.~Mei, X.~Liu, D.~Mandic, W.~Wang, and M.~D.
  Plumbley, ``{{A}udio{LDM}: Text-to-Audio Generation with Latent Diffusion
  Models},'' \emph{Proc. ICML}, 2023.

\bibitem{huang2023makeanaudio}
R.~Huang, J.~Huang, D.~Yang, Y.~Ren, L.~Liu, M.~Li, Z.~Ye, J.~Liu, X.~Yin, and
  Z.~Zhao, ``{{M}ake-{A}n-{A}udio: Text-To-Audio Generation with
  Prompt-Enhanced Diffusion Models},'' in \emph{Proc. ICML}, 2023.

\bibitem{huang2023noise2music}
Q.~Huang, D.~S. Park, T.~Wang, T.~I. Denk, A.~Ly, N.~Chen, Z.~Zhang, Z.~Zhang,
  J.~Yu, C.~Frank, J.~Engel, Q.~V. Le, W.~Chan, Z.~Chen, and W.~Han,
  ``{{N}oise2{M}usic: Text-conditioned Music Generation with Diffusion
  Models},'' \emph{arXiv preprint arXiv:2302.03917}, 2023.

\bibitem{agostinelli2023musiclm}
A.~Agostinelli, T.~I. Denk, Z.~Borsos, J.~Engel, M.~Verzetti, A.~Caillon,
  Q.~Huang, A.~Jansen, A.~Roberts, M.~Tagliasacchi, M.~Sharifi, N.~Zeghidour,
  and C.~Frank, ``{{M}usic{LM}: Generating Music From Text},'' \emph{arXiv
  preprint arXiv:2302.03917}, 2023.

\bibitem{schuhmann2022laion5b}
C.~Schuhmann, R.~Beaumont, R.~Vencu, C.~Gordon, R.~Wightman, M.~Cherti,
  T.~Coombes, A.~Katta, C.~Mullis, M.~Wortsman, P.~Schramowski, S.~Kundurthy,
  K.~Crowson, L.~Schmidt, R.~Kaczmarczyk, and J.~Jitsev, ``{LAION-5B: An open
  large-scale dataset for training next generation image-text models},'' in
  \emph{NeurIPS 2022 Datasets and Benchmarks}, 2022.

\bibitem{lee2023bigvgan}
S.~Lee, W.~Ping, B.~Ginsburg, B.~Catanzaro, and S.~Yoon, ``{Big{VGAN}: A
  Universal Neural Vocoder with Large-Scale Training},'' in \emph{Proc. ICLR},
  2023.

\bibitem{nichol2021ddpm}
A.~Nichol and P.~Dhariwal, ``{Improved Denoising Diffusion Probabilistic
  Models},'' in \emph{Proc. ICML}, 2019.

\bibitem{radford2021clip}
A.~Radford, J.~W. Kim, C.~Hallacy, A.~Ramesh, G.~Goh, S.~Agarwal, G.~Sastry,
  A.~Askell, P.~Mishkin, J.~Clark, G.~Krueger, and I.~Sutskever, ``{Learning
  Transferable Visual Models From Natural Language Supervision},'' in
  \emph{Proc. ICML}, 2021, pp. 8748--8763.

\bibitem{ramesh2022dalle2}
A.~Ramesh, P.~Dhariwal, A.~Nichol, C.~Chu, and M.~Chen, ``{Hierarchical
  Text-Conditional Image Generation with CLIP Latents},'' \emph{arXiv preprint
  arXiv:2204.06125}, 2022.

\bibitem{owens2016vis}
A.~Owens, P.~Isola, J.~McDermott, A.~Torralba, E.~H. Adelson, and W.~T.
  Freeman, ``{Visually Indicated Sounds},'' in \emph{Proc. CVPR}, 2016, pp.
  2405--2413.

\bibitem{iashin2021specvqgan}
V.~Iashin and E.~Rahtu, ``{Taming Visually Guided Sound Generation},'' in
  \emph{Proc. BMVC}, 2021.

\bibitem{sheffer2022im2wav}
R.~Sheffer and Y.~Adi, ``{I Hear Your True Colors: Image Guided Audio
  Generation},'' in \emph{Proc. ICASSP}, 2023.

\bibitem{dong2023clipsep}
H.-W. Dong, N.~Takahashi, Y.~Mitsufuji, J.~McAuley, and T.~Berg-Kirkpatrick,
  ``{{CLIPS}ep: Learning Text-queried Sound Separation with Noisy Unlabeled
  Videos},'' in \emph{Proc. ICLR}, 2023.

\bibitem{hdong2023clipsynth}
H.-W. Dong, G.~Sigurdsson, C.~Tao, J.-Y. Kao, Y.-H. Lin, A.~Narayan-Chen,
  A.~Gupta, T.~Chung, J.~Huang, N.~Peng, and W.~Zhao, ``{CLIPSynth: Learning
  Text-to-audio Synthesis from Videos Using CLIP and Diffusion Models},'' in
  \emph{CVPR Workshop on Sight and Sound}, 2023.

\bibitem{kong2021diffwave}
Z.~Kong, W.~Ping, J.~Huang, K.~Zhao, and B.~Catanzaro, ``{Diff{W}ave: A
  Versatile Diffusion Model for Audio Synthesis},'' in \emph{Proc. ICLR}, 2021.

\bibitem{pascual2022diffusion}
S.~Pascual, G.~Bhattacharya, C.~Yeh, J.~Pons, and J.~Serrà, ``{Full-band
  General Audio Synthesis with Score-based Diffusion},'' in \emph{Proc.
  ICASSP}, 2023.

\bibitem{ho2021classifierfree}
T.~S. Jonathan~Ho, ``{Classifier-Free Diffusion Guidance},'' in \emph{NeurIPS
  Workshop on Deep Generative Models and Downstream Applications}, 2021.

\bibitem{chen2020vggsound}
H.~Chen, W.~Xie, A.~Vedaldi, and A.~Zisserman, ``{{VGGS}ound: A Large-scale
  Audio-Visual Dataset},'' in \emph{Proc. ICASSP}, 2020, pp. 721--725.

\bibitem{zhao2018sop}
H.~Zhao, C.~Gan, A.~Rouditchenko, C.~Vondrick, J.~McDermott, and A.~Torralba,
  ``{The Sound of Pixels},'' in \emph{Proc. ECCV}, 2018.

\bibitem{improved_diff_github}
\url{https://github.com/openai/improved-diffusion}.

\bibitem{clip_github}
\url{https://github.com/openai/CLIP}.

\bibitem{dalle2_huggingface}
\url{https://huggingface.co/laion/DALLE2-PyTorch}.

\bibitem{dalle2_github}
\url{https://github.com/lucidrains/DALLE2-pytorch}.

\bibitem{clap_github}
\url{https://github.com/LAION-AI/CLAP}.

\bibitem{bigvgan_github}
\url{https://github.com/NVIDIA/BigVGAN}.

\bibitem{kilgour2019fad}
K.~Kilgour, M.~Zuluaga, D.~Roblek, and M.~Sharifi, ``{Fréchet Audio Distance:
  A Metric for Evaluating Music Enhancement Algorithms},'' in \emph{Proc.
  INTERSPEECH}, 2019, pp. 2350--2354.

\bibitem{fad_github}
\url{https://github.com/gudgud96/frechet-audio-distance}.

\bibitem{hershey2017cnn}
S.~Hershey, S.~Chaudhuri, D.~P. Ellis, J.~F. Gemmeke, A.~Jansen, R.~C. Moore,
  M.~Plakal, D.~Platt, R.~A. Saurous, B.~Seybold, \emph{et~al.}, ``{CNN
  Architectures for Large-scale Audio Classification},'' in \emph{Proc.
  ICASSP}, 2017, pp. 131--135.

\bibitem{wu2022wav2clip}
H.-H. Wu, P.~Seetharaman, K.~Kumar, and J.~P. Bello, ``{{W}av2{CLIP}: Learning
  Robust Audio Representations From {CLIP}},'' in \emph{Proc. ICASSP}, 2022,
  pp. 4563--4567.

\bibitem{rouditchenko2021avlnet}
A.~Rouditchenko, A.~Boggust, D.~Harwath, B.~Chen, D.~Joshi, S.~Thomas,
  K.~Audhkhasi, H.~Kuehne, R.~Panda, R.~Feris, B.~Kingsbury, M.~Picheny,
  A.~Torralba, and J.~Glass, ``{{AVL}net: Learning Audio-Visual Language
  Representations from Instructional Videos},'' in \emph{Proc. INTERSPEECH},
  2021, pp. 1584--1588.

\end{thebibliography}

\appendix

\section{Implementation Details of the Diffusion Prior Models}
\label{sec:detail_prior}

The diffusion prior models used in this paper are based on the open-source implementation of DALL-E 2 in~\cite{dalle2_github}. Specifically, the input to the models is a sequence formed in the order of the encoded CLIP text tokens, the CLIP text embedding, the diffusion step embedding, the noised CLIP image embedding, and a learnable final input embedding. This sequence is fed to a 12-layer transformer consisting of causal multi-head self-attention and feed-forward networks. The last layer's final output vector corresponding to the final input embedding serves as the prediction of the target CLIP image embedding.

For the diffusion prior model used in CLIPSonic-SD, we use a cosine noise schedule with 1000 diffusion steps during training, and 64 steps at inference time. At each diffusion step during training, we minimize the mean squared error between the predicted and the target CLIP image embeddings. Based on DALL-E 2~\cite{ramesh2022dalle2}, we also explore the classifier-free guidance for training the diffusion prior models by randomly replacing the encoded text tokens and the CLIP text embedding with learnable placeholders 10\% of the time. However, at inference time, we empirically find that using no guidance yields the best results. At inference time, for each CLIP text embedding, we generate two CLIP image embeddings from the diffusion prior model, and select the one with a higher cosine similarity to the CLIP text embedding. To train the model, we use the AdamW optimizer with a learning rate of 0.0001, a batch size of 32, a weight decay of 0.06, and we apply an
exponential moving average on the model parameters with a decay factor of 0.9999. The diffusion prior models in CLIPSonic-SD are trained on MUSIC and VGGSound independently
until convergence at around 200\,k steps.

\section{CLAP Scores for BigVGAN Reconstructions}
\label{sec:bigvgan_clap}

In \cref{fig:results}, we observe that the CLAP scores of the BigVGAN reconstruction using the ground truth mel spectrogram, in many cases, are lower than those of the proposed systems, which indicates lower relevance between the ground truth audio and the text query. In order to adhere to the length of the test data, the BigVGAN CLAP scores are obtained based on the entire 10-sec audio samples. However, empirical listening finds that some segments within the 10-sec samples correspond poorly to the text queries. To further investigate the correspondence, We also compute the BigVGAN CLAP scores using a 4-sec sliding window (consistent with the synthesized sample length) with a hop size of 0.5 sec, and report the maximum, mean, and the minimum scores over all 4-sec segments within a 10-sec sample as the overall score of that sample.

As shown in \cref{tab:slide_clap}, the maximum scores on both datasets are higher than the rest, which supports our observation by listening. On VGGSound, the maximum CLAP score also exceeds those of CLIPSonic-ZS, CLIPSonic-PD, and CLIPSonic-SD (see \cref{tab:results}). On MUSIC, there is a smaller gap between the maximum CLAP score and that obtained using the entire 10-sec audio, indicating a more uniform relevance level within a sample. However, the studied models trained on MUSIC still outperform the BigVGAN reconstruction in terms of the maximum CLAP score (except for CLIPSonic-PD, and CLIPSonic-ZS without using the classifier free guidance, see \cref{fig:results}). In addition to the contribution of the classifier free guidance (\cref{sec:obj}), the remaining reason requires further investigation. Possible directions include manually inspecting and removing samples with poor audio-text correspondence, and also finetuning CLAP on MUSIC.

\begin{table}
    \centering
    \vspace{-1.5ex}
    \caption{CLAP scores computed on BigVGAN reconstructions using a sliding window.}
    \vspace{1ex}
    \begin{tabular}{ccc@{~~~}c}
        \toprule
        Window size &Mode &VGGSound &MUSIC \\ 
        \midrule
        4 sec       &Max  &0.273    &0.280\\
        4 sec       &Mean &0.195    &0.234\\
        4 sec       &Min  &0.111    &0.185\\
        \cmidrule(lr){1-4}
        10 sec      &-    &0.204    &0.272\\
        \bottomrule
    \end{tabular}
    \label{tab:slide_clap}
\end{table}


\section{Limitations}
\label{sec:limitations}

We observe some limitations of the proposed method. First, as CLIPSonic is conditioned on the CLIP embedding of a single video frame, it is not readily applicable to handle more complex text queries that involve sequences of events or dynamic interactions between objects. A more powerful language-vision model that can understand videos is required to apply our proposed method to leverage the rich temporal information in videos. Second, since the conditioning signals are extracted from videos, CLIPSonic cannot learn audio concepts that have little meaning in the visual domain, such as pitch, prosody, genre, and tempo. This represents one of the fundamental limitations of approaches that use the visual domain as a bridge to learn the text-audio correspondence. Finally, CLIPSonic offers limited controllability in generating semantically complex audio, such as speech or music given specific words or scores, respectively. However, the proposed method may serve as a pretraining approach for training language-audio models, where we can first pretrain a language-audio model on a large dataset with only unlabeled videos and later finetune the model on a small dataset with audio-text pairs.

\end{document}